\documentclass[12pt]{article}

\usepackage{a4wide}
\usepackage[pdftex,usenames,dvipsnames]{color}
\usepackage{graphics}
\usepackage{amsfonts}
\usepackage{amssymb}
\usepackage{accents}

\newcommand{\nit}{\noindent}

\newcommand{\np}{\newpage}
\newcommand{\dsp}{\displaystyle}
\newcommand{\vs}[1]{\vspace{#1 ex}}
\newcommand{\hs}[1]{\hspace{#1 em}}
\newcommand{\bfr}{\begin{flushright}}
\newcommand{\efr}{\end{flushright}}
\newcommand{\bc}{\begin{center}}
\newcommand{\ec}{\end{center}}
\newcommand{\ben}{\begin{enumerate}}
\newcommand{\een}{\end{enumerate}}

\newcommand{\be}{\begin{equation}}
\newcommand{\ee}{\end{equation}}
\newcommand{\ba}{\begin{array}}
\newcommand{\ea}{\end{array}}
\newcommand{\ct}{\cite}
\newcommand{\bit}{\bibitem}
\newcommand{\dd}[2]{\frac{\partial{#1}}{\partial{#2}}}
\newcommand{\ag}{\alpha}
\newcommand{\bg}{\beta}

\newcommand{\del}{\delta}

\newcommand{\kg}{\kappa}
\newcommand{\lb}{\lambda}
\newcommand{\sg}{\sigma}
\newcommand{\rg}{\rho}

\newcommand{\Gam}{\Gamma}

\newcommand{\Sg}{\Sigma}

\newcommand{\lh}{\left(}
\newcommand{\rh}{\right)}

\newcommand{\nb}{\nabla}

\newcommand{\der}{\partial}

\begin{document}

\pagestyle{empty}

\bc
\vs{5}
{\large \bf Classical dynamics of particles with non-abelian gauge charges} \\
\vs{5}

{\large Jan W. van Holten} \\
\vs{3}

Nikhef, Amsterdam NL \\
\vs{2}

and \\
\vs{2}

Lorentz Institute, Leiden University \\
\vs{2}

Leiden, NL \\
\vs{3}

January 16, 2025
\ec
\vs{5}

\nit
{\small 
{\bf Abstract} \\
The classical dynamics of particles with (non-)abelian charges and spin moving on curved 
manifolds is established in the Poisson-Hamilton framework. Equations of motion are 
derived for the minimal quadratic Hamiltonian and some extensions involving 
spin-dependent interactions. It is shown that these equations of motion coincide with the 
consistency conditions for current- and energy-momentum conservation. The classical
equations cannot be derived from an action principle without extending the model.
One way to overcome this problem is the introduction of anticommuting Grassmann 
co-ordinates. A systematic derivation of constants of motion based on symmetries of 
the background fields is presented. 
}

\np
\pagestyle{plain} 
\pagenumbering{arabic}

\nit
{\bf 1.\ Poisson-Hamilton dynamics of point particles}
\vs{1}

\nit
The standard classical dynamics of point particles with non-abelian charges was formulated 
in early papers by Kerner \ct{kerner:1968}  and Wong \ct{wong:1970}. In these papers the 
equations of motion were obtained from field theoretical models: geodesic motion in a 
Kaluza-Klein model and the classical limit of the Dirac-Yang-Mills equations, respectively. 
Later authors have elaborated on the geometric derivation in the context of fibre bundle
constructions \ct{cho:1975}-\ct{storchak:2014}. A historical overview is given 
in \ct{horvathy:2023}. In this paper I present a different derivation, self-contained, in a 
hamiltonian framework that allows generalizing the theory to gravitating particles with 
spin and extend it to include spin-dependent interactions such as those of the Stern-Gerlach 
type. 

The point particles considered in this paper are characterized by a worldline $\xi^{\mu}(\tau)$, a
curve on a smooth manifold with metric $g_{\mu\nu}(x)$; their dynamics is formulated in terms of a 
set of phase space variables consisting of the position co-ordinates $\xi^a$, the covariant 
momentum components $\pi_{\mu}$, non-abelian internal momenta $\sg_i$ and spin degrees 
of freedom combined in the anti-symmetric spin tensor $\Sg^{\mu\nu}$ 
\ct{vholten:1993,vholten:2006,vholten:2015}. The starting point for their dynamical 
equations is the set of Poisson brackets
\be
\ba{l}
\dsp{ \left\{ \xi^{\mu}, \xi^{\nu} \right\} = 0, \hs{1.3} \left\{ \xi^{\mu}, \pi_{\nu} \right\} = \del_\nu^\mu\, 
\hs{4} \left\{ \pi_{\mu}, \pi_{\nu} \right\} = g_{\sg} \sg_i F_{\mu\nu}^i + 
\frac{1}{2}\, \Sg^{\kg\lb} R_{\kg\lb\mu\nu}, }\\
 \\
\dsp{ \left\{ \xi^{\mu}, \sg_i \right\} = 0, \hs{1.5} 
 \left\{ \pi_{\mu}, \sg_i \right\} = g_{\sg} f_{ij}^{\;\;\,k} A_{\mu}^j \sg_k, 
 \hs{0.6} \left\{ \sg_i, \sg_j \right\} = f_{ij}^{\;\;\,k} \sg_k, }\\
 \\
\dsp{ \left\{ x^{\mu}, \Sg^{\nu\lb} \right\} = 0, \hs{1} 
 \left\{ \pi_{\mu}, \Sg^{\nu\lb} \right\} = \Gam_{\mu\kg}^{\;\;\;\nu} \Sg^{\kg\lb}
 - \Gam_{\mu\kg}^{\;\;\;\lb} \Sg^{\kg\nu}, }\\ 
 \\
\dsp{ \left\{ \Sg^{\mu\nu}, \Sg^{\kg\lb} \right\} = g^{\mu\kg} \Sg^{\nu\lb} - g^{\mu\lb} \Sg^{\nu\kg} - 
 g^{\nu\kg} \Sg^{\mu\lb} + g^{\nu\lb} \Sg^{\mu\kg}. }
\ea
\label{1}
\ee
These brackets define a consistent, closed Poisson algebra, satisfying all Jacobi identities 
independent of any specific hamiltonian. Note that the internal momenta define a (usually 
compact) closed Lie algebra with structure constants $f_{ij}^{\;\;\,k}$, and the components 
of the spin tensor define a closed lorentzian Poisson subalgebra. The Yang-Mills fields, 
space-time metric, connection and Riemann tensor appear as structure functions of the 
algebra; $g_{\sg}$ is the Yang-Mills coupling constant, to be distinguished from the 
determinant of the metric $g = \det g_{\mu\nu}$.

The standard dynamics of a massive particle in the presence of external gauge- and 
gravitational fields is obtained taking Poisson brackets of observables $J(\xi, \pi, \sg, \Sg)$ 
with the minimal proper-time hamiltonian 
\be
H_m = \frac{1}{2m}\, g^{\mu\nu} \pi_{\mu} \pi_{\nu} \hs{1} \mbox{and} \hs{1} 
\dot{J} = \left\{ J, H_m \right\}.
\label{2}
\ee
With this choice of hamiltonian the equations of motion become
\be
\ba{l}
m\, \dot{\xi}^{\mu} = g^{\mu\nu} \pi_{\nu}, \\
 \\
\dsp{ \frac{D\pi_{\mu}}{D\tau} = m g_{\mu\nu} \lh \ddot{\xi}^{\nu} + 
 \Gam_{\kg\lb}^{\;\;\;\nu} \dot{\xi}^{\kg} \dot{\xi}^{\lb} \rh 
 = g_{\sg} \sg_i F_{\mu\nu}^i\, \dot{\xi}^{\nu} + \frac{1}{2}\, \Sg^{\kg\lb} R_{\kg\lb\mu\nu}\, \dot{\xi}^{\nu}, }
\ea
\label{3}
\ee
\[
\ba{lr}
\dsp{ \frac{D\sg_i}{D\tau} = \dot{\sg}_i + g_{\sg} f_{ij}^{\;\;\;k} \dot{\xi}^{\mu} A^j_{\mu} \sg_k = 0, } & \\
 \\
\dsp{ \frac{D \Sg^{\mu\nu}}{D\tau} = \dot{\Sg}^{\mu\nu} + 
 \dot{\xi}^{\kg} \lh \Gam_{\kg\lb}^{\;\;\;\mu} \Sg^{\lb\nu} 
 - \Gam_{\kg\lb}^{\;\;\;\nu} \Sg^{\lb\mu} \rh = 0. } & \hs{5} (3)\; \mbox{(cont'd)}
\ea
\]
Thus the internal momenta and spin tensor are covariantly constant, whilst the 
world line deviates from a geodesic because of the non-abelian Lorentz force and its
spin-dependent gravitational analogue. It follows that with this choice of hamiltonian 
the total internal momentum $\sg_i \sg^i$ and the total spin $\Sg^{\mu\nu} \Sg_{\mu\nu}$ 
are constant, as a result of which they can only perform precessional motion. 

For any two scalar functions on the phase space $J$, $K$ which can be an expanded as
\be
J(x,\pi,\sg,\Sg) = \sum_{n\geq 0} \frac{1}{n!}\, J^{(n)\,\mu_1 ..\mu_n}(x,\sg,\Sg) \pi_{\mu_1} .. \pi_{\mu_n}
\label{2.a}
\ee
the Poisson bracket takes the form
\be
\ba{lll}
\left\{ J, K \right\} & = & \dsp{ D_{\mu} J\, \dd{K}{\pi_{\mu}} - \dd{J}{\pi_{\mu}}\, D_{\mu} K + 
 \lh g_{\sg} \sg_i  F^i_{\mu\nu} + \frac{1}{2}\, \Sg^{\kg\lb} R_{\kg\lb\mu\nu} \rh \dd{J}{\pi_{\mu}} 
 \dd{K}{\pi_{\nu}} }\\
 & & \\
 & & \dsp{ +\, f_{ij}^{\;\;\,k} \sg_k\, \dd{J}{\sg_i} \dd{K}{\sg_j} + 
 4 g^{\mu\kg} \Sg^{\nu\lb}\, \dd{J}{\Sg^{\mu\nu}} \dd{K}{\Sg^{\kg\lb}} }
\ea
\label{1.a}
\ee
in which $D_{\mu}$ is a form of general and gauge covariant derivative \ct{vholten:2006}
\be
D_{\mu} J = \dd{J}{\xi^{\mu}} + \Gam_{\mu\nu}^{\;\;\;\lb} \pi_{\lb} \dd{J}{\pi_{\nu}} - 
 2 \Gam_{\mu\lb}^{\;\;\;\nu}\, \Sg^{\lb\kg} \dd{J}{\Sg^{\nu\kg}} - 
 g_{\sg} f_{ij}^{\;\;\,k} A^j_{\mu} \sg_k \dd{J}{\sg_i}.
\label{1.b}
\ee
With the minimal hamiltonian (\ref{2}) this indeed leads to the evolution equation
\be
\dot{J} = \left\{ J, H_m \right\} = \dot{\xi}^{\mu} \left[ D_{\mu} J - \lh g_{\sg} \sg_i F^i_{\mu\nu} + 
 \frac{1}{2}\, \Sg^{\kg\lb} R_{\kg\lb\mu\nu} \rh \dd{J}{\pi_{\nu}} \right].
\label{2.b}
\ee

\nit
{\bf 2.\ Current and energy-momentum conservation}
\vs{1}

\nit
The inhomogeneous Yang-Mills and Einstein field equations 
\be
\nb_{\mu} F_i^{\mu\nu} = j_i^{\nu}, \hs{2} R_{\mu\nu} - \frac{1}{2}\, g_{\mu\nu} R = - 8\pi G T_{\mu\nu}, 
\label{4}
\ee
require for consistency the gauge- and reparametrization-covariant conservation conditions
\be
\ba{l}
\dsp{ \nb_{\mu} j_i^{\mu} = \der_{\mu}  j_i^{\mu} + g_{\sg} f_{ij}^{\;\;\;k} A^j_{\mu} j_k^{\mu} = 0, }\\
 \\
\dsp{ \nb_{\mu} T^{\mu\nu} = \lh \der_{\mu} T^{\mu\nu} + \Gam_{\mu\lb}^{\;\;\;\mu} T^{\lb\nu} 
 + \Gam_{\mu\lb}^{\;\;\;\nu} T^{\mu\lb} \rh = 0. }
\ea
\label{5}
\ee
These conditions are satisfied for point sources with world line $\xi^{\mu}(\tau)$ satisfying 
the equations (\ref{3}) when the current density vector and energy-momentum tensor 
are given by the proper-time integrals\footnote{Here $\del^4(x-y)$ is a scalar distribution
such that $\int d^4x \sqrt{-g}\, f(x) \del^4(x-y) = f(y)$. }
\be
j_i^{\mu} = g_{\sg} \int d\tau\, \sg_i(\tau)\, \dot{\xi}^{\mu}(\tau) \del^4\lh x - \xi(\tau) \rh,
\label{6}
\ee
and 
\[
T^{\mu\nu} = T_m^{\mu\nu} + T_{YM}^{\mu\nu}, 
\]
where
\be
\ba{lll}
T^{\mu\nu}_m & = & \dsp{ m \int d\tau\, \dot{\xi}^{\mu} \dot{\xi}^{\nu}\, \del^4\lh x - \xi(\tau) \rh }\\
 & & \\
 & & \dsp{ +\, \frac{1}{2}\, \nb_{\lb} \int d\tau \lh \dot{\xi}^{\mu} \Sg^{\nu\lb} + 
 \dot{\xi}^{\nu} \Sg^{\mu\lb} \rh \del^4\lh x - \xi(\tau) \rh, }\\
 & & \\
T_{YM}^{\mu\nu} & = & \dsp{  F_i^{\mu\lb} F^{i\nu}_{\;\;\,\lb} - 
 \frac{1}{4}\, g^{\mu\nu} F_i^{\kg\lb} F^i_{\kg\lb}. }
\ea
\label{7}
\ee

\nit
{\bf 3.\ Spin-dependent interactions} 
\vs{1}

\nit
Using the Poisson algebra (\ref{1}) more general equations of motion can be derived using 
additional interaction terms in the hamiltonian. For example, spin-dependent Stern-Gerlach 
type interactions can be modeled by taking 
\be
H = H_m + \ag H_1 + \bg H_2
\label{8}
\ee
with 
\be
H_1 = \frac{1}{2}\,\sg_i F^i_{\mu\nu} \Sg^{\mu\nu} \hs{1} \mbox{and} \hs{1} 
H_2 = \frac{1}{4}\, R_{\mu\nu\kg\lb} \Sg^{\mu\nu} \Sg^{\kg\lb}.
\label{9}
\ee
The equations of motion (\ref{3}) are now extended to 
\be
\ba{lll}
m\, \dot{\xi}^{\mu} & = & g^{\mu\nu} \pi_{\nu}, \\
 \\
\dsp{ \frac{D\pi_{\mu}}{D\tau} }& = & \dsp{ g_{\sg} \sg_i F_{\mu\nu}^i\, \dot{\xi}^{\nu} + 
 \frac{1}{2}\, \Sg^{\kg\lb} R_{\kg\lb\mu\nu}\, \dot{\xi}^{\nu} - 
 \frac{\ag}{2}\, \sg_i \Sg^{\nu\lb} \nb_{\mu} F^i_{\nu\lb} - 
 \frac{\bg}{4}\, \Sg^{\nu\kg} \Sg^{\lb\rg} \nb_{\mu} R_{\nu\kg\lb\rg}, }\\
\\
\dsp{ \frac{D\sg_i}{D\tau} }& = & \dsp{ \frac{\ag}{2}\, f_{ij}^{\;\;\,k} \sg_k \Sg^{\mu\nu} F_{\mu\nu}^j, }\\
 \\
\dsp{ \frac{D \Sg^{\mu\nu}}{D\tau} }& = & \dsp{ - \ag \sg_i \lh F^{i\mu}_{\;\;\lb} \Sg^{\lb\nu} - 
 F^{i\nu}_{\;\;\lb} \Sg^{\lb\mu} \rh
 + \bg\, \Sg^{\kg\lb} \lh R_{\kg\lb\rg}^{\;\;\;\;\;\mu} \Sg^{\rg\nu} - R_{\kg\lb\rg}^{\;\;\;\;\;\nu} \Sg^{\rg\mu} \rh. }
\ea
\label{10}
\ee
These equations are still consistent with the conservation of total internal momentum $\sg_i \sg^i$ 
and spin $\Sg^{\mu\nu} \Sg_{\mu\nu}$. They also remain consistent with current and 
energy-momentum conservation at the price of extending the expressions for the current density 
vector and the energy-momentum tensor to 
\be
j_i^{\mu} = g_{\sg} \int d\tau\, \sg_i \dot{\xi}^{\mu} \del^4\lh x - \xi(\tau) \rh - \ag \nb_{\nu}
 \int d\tau\, \sg_i \Sg^{\mu\nu} \del^4\lh x - \xi(\tau) \rh,  
\label{11}
\ee
and
\[
T^{\mu\nu} = T_m^{\mu\nu} + T_{YM}^{\mu\nu} + \ag T_1^{\mu\nu} + \bg\, T_2^{\mu\nu}, 
\]
with 
\be
\ba{lll}
T_1^{\mu\nu} & = & \dsp{  \frac{1}{2}\,  \int d\tau\, \sg_i \lh F^{i\mu}_{\;\;\lb} \Sg^{\lb\nu} + 
 F^{i\nu}_{\;\;\lb}\Sg^{\lb\mu} \rh \del^4\lh x - \xi(\tau) \rh }\\
 & & \\
T_2^{\mu\nu} & = & \dsp{ \frac{1}{4}\, \int d\tau\, \Sg^{\kg\lb} \lh R_{\kg\lb\rg}^{\;\;\;\;\;\,\nu}\Sg^{\rg\mu} 
 + R_{\kg\lb\rg}^{\;\;\;\;\;\,\mu}\Sg^{\rg\nu} \rh  \del^4\lh x - \xi(\tau) \rh }\\
 & & \\
 & & \dsp{ +\, \frac{1}{2}\, \nb_{\kg} \nb_{\lb} \int d\tau \lh \Sg^{\mu\kg} \Sg^{\lb\nu} 
 + \Sg^{\nu\kg} \Sg^{\lb\mu} \rh \del^4\lh x - \xi(\tau) \rh. }
\ea
\label{12}
\ee

\nit
{\bf 4.\ Actions}
\vs{1}

\nit
Remarkably, the {Poisson-Hamilton equations of motion derived above cannot be obtained 
from an action principle, at least not without modifying the phase space with additional or 
different degrees of freedom \ct{horvathy:1982,balachandran:2017}. A clear indication of 
this result is that the non-abelian internal momenta and spin tensor do not have conjugate 
variables, which are necessary to produce equations of motion which are first-order 
differential equations in proper time. 

Horvathy pointed out that an action can be defined by introducing co-ordinates on the 
gauge-group manifold as additional degrees of freedom 
\ct{horvathy:1982,balachandran:2017}.\footnote{A variant on this derivation
using a many-particle formulation in the infinite-momentum frame is presented in 
\cite{venugopalan:2001}.} Another way out is to represent the internal and spin degrees 
of freedom by Grassmann variables instead. Indeed, the internal momenta can be replaced 
by Grassmann variables $\eta^i$ such that 
\be
\sg_i = \frac{i}{2}\, f_{ijk} \eta^j \eta^k, 
\label{13}
\ee
and introduce the graded Poisson brackets 
\be
\left\{ \eta^i, \eta^j \right\} = i \del^{ij}, \hs{2} 
\left\{ \pi_{\mu}, \eta^i \right\} = g_{\sg} f_{jk}^{\;\;\,i} A_{\mu}^j \eta^k.
\label{14}
\ee
These relations reproduce all the brackets involving $\sg_i$ in eq.\ (\ref{1}). The equation
of motion for the $\eta^i$ with the standard hamiltonian now becomes
\be
\frac{D\eta^i}{D\tau} = \dot{\eta}^i + g_{\sg} f_{jk}^{\;\;\,i} \dot{\xi}^{\mu} A^j_{\mu} \eta^k = 0 \hs{1} 
\Rightarrow \hs{1} \frac{D\sg_i}{D\tau} = 0.
\label{15}
\ee
Similarly we can represent the spin degrees of freedom by Grassmann variables $\psi^a$ 
taking \ct{berezin:1977}
\be
\Sg^{\mu\nu} = i \psi^{\mu} \psi^{\nu},
\label{16}
\ee
with the graded brackets 
\be
\left\{ \psi^{\mu}, \psi^{\nu} \right\} = i g^{\mu\nu}, \hs{2} 
\left\{ \pi_{\mu}, \psi^{\nu} \right\} = \Gam_{\mu\lb}^{\;\;\;\nu} \psi^{\lb},
\label{17}
\ee
reproducing all our previous results. With the minimal hamiltonian $H_m$ its
equation of motion is
\be
\frac{D\psi^{\mu}}{D\tau} = \dot{\psi}^{\mu} + \dot{\xi}^{\nu} \Gam_{\nu\lb}^{\;\;\;\mu} \psi^{\lb} = 0 \hs{1} 
\Rightarrow \hs{1} \frac{D\Sg^{\mu\nu}}{D\tau} = 0.
\label{18}
\ee
The equations framed in terms of Grassmann variables can now be derived from an action principle 
using the action 
\be
S_m = \int d\tau\, \lh \dot{x}^{\mu} \pi_{\mu} + \frac{i}{2}\, \del_{ij} \eta^i\, \frac{D\eta^j}{D\tau}  +
 \frac{i}{2}\, g_{\mu\nu} \psi^{\mu} \frac{D\psi^{\nu}}{D\tau} - H_m \rh.
\label{19}
\ee
In consequence of their anticommutation properties the first-order kinetic terms for the 
Grassmann variables do not reduce to a total derivative. It is straightforward to complete 
this action to one with full world-line supersymmetry  \ct{balachandran:2017}, 
\ct{vholten:1993}-\ct{govaerts:2014} by including Stern-Gerlach terms specific to 
Dirac fermions. This requirement follows immediately from the supersymmetry 
algebra with the supercharge $Q = \psi^{\mu} \pi_{\nu}$ as
\be
2m H = - i \left\{ Q, Q \right\} = g^{\mu\nu} \pi_{\mu} \pi_{\nu} - 
 \Sg^{\mu\nu} \lh g_{\sg} \sg_i F^i_{\mu\nu} + \frac{1}{2} \Sg^{\kg\lb} R_{\kg\lb\mu\nu} \rh.
\label{20}
\ee
The quantum version of this model then becomes the Einstein-Dirac-Yang-Mills theory. 
Finally it goes without saying that abelian gauge interactions are included in all of the 
above after simply replacing an internal momentum $\sg_i$ by a constant non-dynamical 
charge $q$ and taking $f_{ij}^{\;\;\,k} = 0$, without need for expressing the charge 
in terms of Grassmann variables. 
\vs{2}

\nit
{\bf 5.\ Constants of motion}
\vs{1}

\nit
With the general framework established above the motion of non-abelian particles can
be studied for specific gravitational and Yang-Mills backgrounds. An important tool for
finding solutions is the identification of constants of motion, arising from generalized 
symmetries of the background. Such constants are solutions of the condition 
\be
\dot{J} = \left\{ J, H \right\} = 0. 
\label{5.1}
\ee
By eq.\ (\ref{2.b}) with the minimal hamiltonian this becomes more explicitly
\be
D_{\mu} J = \lh g_{\sg} \sg_i F^i_{\mu\nu} + \frac{1}{2}\, \Sg^{\kg\lb} R_{\kg\lb\mu\nu} \rh \dd{J}{\pi_{\nu}}.
\label{5.2}
\ee
With the expansion ({\ref{2.a}) this becomes a hierarchy of equations \ct{vholten:2006}
\be
\ba{l}
\dsp{ D_{\mu} J^{(0)} =  \lh g_{\sg} \sg_i F^{i\,\nu}_{\mu} + 
 \frac{1}{2}\, \Sg^{\kg\lb} R_{\kg\lb\mu}^{\;\;\;\;\;\;\nu} \rh J^{(1)}_{\,\nu}, }\\
 \\
\dsp{ D_{\mu} J^{(1)}_{\;\nu} + D_{\nu} J^{(1)}_{\;\mu} = \lh g_{\sg} \sg_i F^{i\;\kg}_{\,\mu} + 
 \frac{1}{2}\, \Sg^{\lb\rg} R_{\lb\rg\mu}^{\;\;\;\;\;\kg} \rh J^{(2)}_{\kg\nu} + \lh g_{\sg} \sg_i F^{i\;\kg}_{\,\nu} 
 + \frac{1}{2}\, \Sg^{\lb\rg} R_{\lb\rg\nu}^{\;\;\;\;\;\kg} \rh J^{(2)}_{\kg\mu} , }
\ea
\label{5.3}
\ee
and at the $n$th level
\be
D_{(mu_1} J^{(n-1)}_{\mu_2...\mu_n)} = \lh g_{\sg} \sg_i F^{i\;\nu}_{(\mu_1} + 
 \frac{1}{2}\, \Sg^{\kg\lb} R_{\kg\lb(\mu_1}^{\;\;\;\;\;\;\nu} \rh J^{(n)}_{\mu_2 .. \mu_n)\nu}, 
\label{5.3a}
\ee
where the parentheses denote complete symmetrization of the indices enclosed. 
This hierarchy can be truncated if the background admits a Killing tensor $J^{(n-1)}(x)$
of rank $n-1$, as by definition this satisfies
\[
D_{(\mu_1} J^{(n-1)}_{\mu_2...\mu_n)}(x) = \nb_{(\mu_1} J^{(n-1)}_{\mu_2...\mu_n)}(x) = 0.
\]
Therefore all component $J^{(p)}$ of rank $p \geq n$ can be taken to vanish, 
and by solving the first $n$ equations (\ref{5.3}) starting with the maximal rank and 
going down to $n = 0$ one obtains the full solution for $J(x,\pi,\sg,\Sg)$. Thus we
have established a one-one correspondence between Killing tensors and those 
constants of motion for models with hamiltonian $H_m$ which are polynomial in 
the momentum components $\pi_{\mu}$. 

For models with hamiltonians which include Stern-Gerlach interactions the full equation 
reads
\be
\ba{lll}
\dot{J} & = & \dsp{ \dot{\xi}^{\mu} \left[ D_{\mu} J - \lh g_{\sg} \sg_i F^i_{\mu\nu} + 
 \frac{1}{2}\, \Sg^{\kg\lb} R_{\kg\lb\mu\nu} \rh \dd{J}{\pi_{\nu}} \right] }\\
 & & \\ 
 & & \dsp{ - \frac{\ag}{2} \left[ \sg_i \Sg^{\nu\lb} \nb_{\mu} F^i_{\nu\lb}\, \dd{J}{\pi_{\mu}} 
 - f_{ij}^{\;\;\,k} \sg_k F^j_{\mu\nu} \Sg^{\mu\nu}\, \dd{J}{\sg_i} + 
 2 \sg_i \lh F^{i\mu}_{\;\;\,\lb} \Sg^{\lb\nu} - F^{i\nu}_{\;\;\,\lb} \Sg^{\lb\mu} \rh \dd{J}{\Sg^{\mu\nu}} \right] }\\
 & & \\
 & & \dsp{ - \frac{\bg}{4} \left[ \Sg^{\nu\kg} \Sg^{\lb\rg} \nb_{\mu} R_{\nu\kg\lb\rg}\, \dd{J}{\pi_{\mu}} - 
 4 \Sg^{\lb\rg}\lh R_{\lb\rg\kg}^{\;\;\;\;\;\;\mu} \Sg^{\kg\nu} - 
 R_{\lb\rg\kg}^{\;\;\;\;\;\;\nu} \Sg^{\kg\mu} \rh \dd{J}{\Sg^{\mu\nu}} \right] = 0. }
\ea
\label{5.4}
\ee
This implies as a direct constraint that the terms on the right-hand side independent 
of $\pi_{\mu}$ must cancel among themselves:
\be
\ba{l}
\dsp{  \Sg^{\mu\nu} \left[ J^{(1)\lb} \nb_{\lb} F^i_{\mu\nu} - 
 f_{jk}^{\;\;\,i}  F^k_{\mu\nu}\, \dd{J^{(0)}}{\sg_j} + 2 \lh F^{i\lb}_{\;\;\,\mu}\, \dd{J^{(0)}}{\Sg^{\lb\nu}} - 
F^{i\lb}_{\;\;\,\nu} \dd{J^{(0)}}{\Sg^{\lb\mu}} \rh \right] = 0 }\\
 \\
\dsp{ \Sg^{\mu\nu} \left[ J^{(1)\kg} \nb_{\kg} R_{\lb\rg\mu\nu} - 
 4 R_{\lb\rg\mu}^{\;\;\;\;\;\kg}\, \dd{J^{(0)}}{\Sg^{\kg\nu}} +  
4 R_{\lb\rg\nu}^{\;\;\;\;\;\kg}\, \dd{J^{(0)}}{\Sg^{\kg\mu}} \right] = 0. }
\ea
\label{5.5}
\ee
As noted before these admit trivial solutions with $J^{(1)\mu} = 0$ in terms of the Casimir 
invariants of the internal symmetry and the lorentzian symmetry
\be
J = (\sg_i\sg^i, \Sg_{\mu\nu} \Sg^{\mu\nu}).
\label{5.6}
\ee
A non-trivial solution of the full equation exists if there is a Killing vector $J^{\mu}$ 
which is an isometry of the metric as well as of the gauge field, as expressed by 
the vanishing of the Lie-derivatives
\[
\ba{l}
J^{\mu} \lh \nb_{\mu} F^i_{\nu\lb} - g f_{jk}^{\;\;\,i} A_{\mu}^j F^k_{\nu\lb} \rh + 
 F^i_{\mu\lb} \nb_\nu J^{\mu} + F^i_{\nu\mu} \nb_{\lb} J^{\mu} = 0, \\
  \\
J^{\mu} \nb_{\mu} R_{\nu\kg\lb\rg} + R_{\mu\kg\lb\rg} \nb_{\nu} J^{\mu} + 
 R_{\nu\mu\lb\rg} \nb_{\kg} J^{\mu} + R_{\nu\kg\mu\rg} \nb_{\lb} J^{\mu} 
 + R_{\nu\kg\lb\mu} \nb_{\rg} J^{\mu} = 0.
\ea
\]
The corresponding solution of equation (\ref{5.4}) then is
\be
J = g_{\sg} \sg_i A^i_{\mu} J^{\mu} + \frac{1}{2}\, \Sg^{\mu\nu} \nb_{\mu} J_{\nu} + J^{\mu} \pi_{\mu}, 
\label{5.7}
\ee
on condition that $\nb_{\mu} J_{\nu} + \nb_{\nu} J_{\mu} = 0$. Solutions with rank $n = 2$ Killing 
tensors appear e.g.\ in supersymmetric models which admit a constant of Runge-Lenz type, as in 
\ct{vholten:1993,vholten:1994}.
\vs{3}

\nit
{\bf Acknowledgement} \\
The author is indebted to Prof.\ P. Horvathy and Prof.\ R. Kerner for critical reading of the 
manuscript. 
\vs{3}

\end{document}